\documentstyle[aps]{revtex}

\title{Effective field theory of resonant $2$-level atoms}

\author{Mark Burgess}

\address{Centre of Science and Technology\\Oslo College\\Cort Adelers Gate 30\\0254 Oslo, Norway}

\def\beq{\begin{eqnarray}}
\def\eeq{\end{eqnarray}}
\def\2{\frac{1}{2}}
\def\32{\frac{\sqrt 3}{2}}
\def\i2{\frac{i}{\sqrt 2}}
\def\12{\frac{1}{\sqrt 2}}

\def\i{\hat{\bf i}}

\begin{document}
\maketitle

\begin{abstract}
The phenomenological two-level atom is re-analysed using the methods of effective
field theory.  By presenting the Dicke-Jaynes-Cummings model in real
space, an exact diagonalization is accomplished going beyond the 
rotating wave approximation. The meaning of the symmetries and
conserved quantities in the theory is discussed and the model is
related to non-equilibrium field theory. The structure of the solution
raises a question about the rotating wave approximation in quantum
mechanics.
\end{abstract}

\vspace{2cm}
It was first realized by Jaynes and Cummings that a semi-classical
model of a two-level atom could reproduce the essential features of
the quantum theoretical problem\cite{jaynes1}.  Recently there has
been a ressurgence of interest in effective field theory in several
works particularly in connection with quantum
electrodynamics\cite{caswell1,labelle1,kong1}. Effective theories are
often the most efficient way of performing calculations in field
theory; they naturally separate conceptually independent issues.  An
alternative to an effective theory is to perform {\em ab initio}
calculations, first principle computations in quantum electrodynamics,
but this is orders of magnitude harder and involves computing effects
in which one is not interested. For instance, if one is not interested
in the reason for the energy spacing of the bound state levels in the
atomic system, it is not neccesary to derive it: rather it can be
inserted either from experiment of other calculations.

This note reexamines an extremely well known and widely used model for
the resonant interaction between a two level atom and radiation.  The
two level system has a broad repertoire of applications in physics,
from spin models to the micromaser\cite{micromaser}.  It is related to
a class of Dicke models\cite{dicke1,gilmore1} and, in the so-called
rotating wave approximation, it becomes the Jaynes-Cummings
model\cite{jaynes1} which may be solved exactly.  A Hamiltonian
analysis of symmetries in this Jaynes-Cummings model is given in
ref. \cite{benivegna1}. In reference
\cite{burgess12} a variation on Schwinger's closed time
path\cite{schwinger2} method of analysing non-equilibrium systems was
presented in which field theories with spacetime dependent
perturbations could be formulated analogously to gauge theories. The
same technique can be applied to the two-level atom to solve the full
model and eliminate the need for the so-called rotating wave
approximation.  In doing so, one obtains extra physical insight
into the system as well as a result which is valid over a wider range of
parameters.  It is also possible to verify a previous claim in
ref. \cite{burgess12} about the relationship between non-equilibrium
effective field theory and Rabi oscillations of the two-level atom.

Consider the phenomenological two-level system described by the action
\beq
S = \int dV_t \left[ 
-\frac{\hbar^2}{2m}(\partial^i\psi_A)^*(\partial_i\psi_A) - \psi^*_A V_{AB}(t)\psi_B
+\frac{i\hbar}{2}(\psi^* D_t \psi - (D_t\psi)^*\psi)
\right]
\eeq
where $A,B=1,2$ characterizes the two levels, $i\hbar D_t =
i\hbar\partial_t + i\Gamma(t)$ in matrix notation, and $\Gamma=\Gamma_{AB}$
is an off-diagonal anti-symmtrical matrix.  At frequencies which are
small compared to the light-size of the atom, an atoms may be
considered electrically neutral. The distribution of charge within the
atoms is not required here. In this approximation the
leading interaction is a resonant dipole transition.  The connection
$\Gamma_{AB}$ plays an analogous role to the electromagnetic vector
potential in electrodynamics, but it possesses no dynamics of its
own. Rather is works as a constraint variable, or auxiliary Lagrange
mulitplier field. There is no electromagnetic vector potential in the
action since the field is electrically neutral in this
formulation. $\Gamma_{AB}$ refers not to the $U(1)$ phase symmetry but
to the two level symmetry.
Variation of the action with respect to
$\Gamma(t)$ provides us with the conserved current.
\beq
\frac{\delta S}{\delta \Gamma_{AB}} = \frac{i}{2} (\psi_A^*\psi_B-\psi^*_B\psi_A)
\eeq
which represents the amplitude for stimulated transition between the
levels.  The current generated by this connection is conserved only on
average, since we are not taking into account any backreaction. The
conservation law corresponds merely to
\begin{equation}
\partial_t \left(\frac{\delta S}{\delta \Gamma_{AB}} \right) \propto \sin(2\int X(t))
\end{equation}
where $X(t)$ will be defined later.
The potential $V_{AB}(t)$ is time dependent and comprises the effect of the
level splitting as well as a perturbation mediated by the radiation
field. A `connection' $\Gamma_{21} = -\Gamma_{12}$ is introduced since
the diagonalization procedure requires a time-dependent unitary
transformation and thus general covariance demands that this will
transform in a different basis. The physics of the model depends on
the initial value of this `connection' and this is the key to the
trivial solubility of the Jaynes-Cummings model.

In matrix form we may write the action for the matter fields
\beq
S = \int dV_t \; \psi^*_A {\cal O}_{AB} \psi_B
\eeq
where
\beq
{\cal O}=\left[ 
\begin{array}{cc}
-\frac{\hbar^2\nabla^2}{2m}-V_1 -\frac{i\hbar}{2}\hbar D_t & J(t)+i\Gamma_{12}\\
J(t)-i\Gamma_{12} &-\frac{\hbar^2\nabla^2}{2m}-V_2 -\frac{i\hbar}{2}\stackrel{\leftrightarrow}{D_t}\\
\end{array}
\right].
\eeq
The level potentials may be regarded as constants in the effective
theory. They are given by $V_1 = E_1$ and $V_2 = E_2 - \hbar\Omega_R$
where $\hbar\Omega_R$ is the interaction energy imparted by the photon
during the transition i.e. the continuous radiation pressure on the
atom. In the effective theory we must add this by hand since we have
separated the levels into independent fields which are electrically
neutral; it would follow automatically in a complete microscopic
theory.  The quantum content of this model is now that this recoil
energy is a quantized unit of $\hbar \Omega$, the energy of a photon
at the frequency of the source. Also the amplitude of the source $J$
would be quantized and proportional to the number of photons on the
field.
If one switches off the source (which models the photon's electric
field) this radiation energy does not automatically go to zero, so
this form is applicable mainly to continuous operation (stimulation).
The origin of the recoil is clear however: it is the electromagnetic
force's interaction with the electron, transmitted to the nucleus by
binding forces. What we are approximating is clearly a $J^\mu A_\mu$
term for the electron, with neutralizing background charge.

It is now desirable to perform a unitary transformation on the action
$\psi \rightarrow U\psi$, ${\cal O} \rightarrow U{\cal O}U^{-1}$
which diagonalizes the operator $\cal O$.  Clearly the connection
 $\Gamma_{AB}$ will transform under this procedure by
\beq
\Gamma \rightarrow \Gamma + \frac{i\hbar}{2} \left( U(\partial_t U^{-1}) - (\partial_t U)U^{-1} \right)
\eeq
since a time-dependent transformation is required to effect the diagonalization.
For notational simplicity we define $\hat L = -\frac{\hbar^2\nabla^2}{2m} -\frac{i}{2}\hbar\stackrel{\leftrightarrow}{D_t}$, so that the secular equation for the
action is:
\beq
(\hat L - E_1-\lambda)(\hat L - E_2 +\hbar\Omega -\lambda) - (J^2+\Gamma_{12}^2) = 0.
\eeq
Note that since $J\stackrel{\leftrightarrow}{\partial_t} J = 0$
there are no operator difficulties with this equation.
The eigenvalues are thus
\beq
\lambda_\pm &=& \hat L - \overline E_{12} + \hbar \Omega 
\pm \sqrt{\frac{1}{4}(\tilde E_{21}-\hbar\Omega)^2 + J^2+\Gamma_{12}^2}\\
&\equiv& \hat L - \overline E_{12} + \hbar \Omega 
\pm \sqrt{\hbar^2\tilde\omega^2 + J^2+\Gamma_{12}^2}\\
&\equiv&  \hat L - \overline E_{12} + \hbar \Omega \pm \hbar \omega_R,
\eeq
where $\overline E_{12} = \2(E_1+E_2)$ and $\tilde
E_{21}=(E_2-E_1)$.  For notational simplicity we define
$\tilde\omega$ and $\omega_R$. One may now confirm this procedure by looking
for the eigenvectors and constructing $U^{-1}$ as the matrix of these
eigenvectors. This may be written in the form
\beq
U^{-1} = \left( 
\begin{array}{cc}
\cos\theta & -\sin\theta\\
\sin\theta & \cos \theta
\end{array}
\right)
\eeq
where
\beq
\cos \theta &=& \frac{\hbar(\tilde \omega+\omega_R)}{\sqrt{\hbar^2(\tilde \omega+\omega_R)^2+J^2+\Gamma_{12}^2}}\\
\sin \theta &=& \frac{\sqrt{J^2+\Gamma^2_{12}}}{\sqrt{\hbar^2(\tilde \omega+\omega_R)^2+J^2+\Gamma_{12}^2}}.
\eeq
The change in the connection $\Gamma(t)$ is thus off-diagonal and anti-symmetric
as required by the gauge symmetry conservation law:
\beq
U\partial_t U^{-1} = \left(
\begin{array}{cc}
0 & \partial_t\theta \\
-\partial_t \theta & 0
\end{array}
\right)
\eeq
The time derivative of $\theta(t)$ may be written in one of two forms which
must agree
\beq
(\partial_t \theta) = \frac{\partial_t\cos\theta}{-\sin\theta}= \frac{\partial_t\sin\theta}{\cos\theta}.
\eeq
This provides a consistency condition which may be verified and leads to the
proof of the identities
\begin{eqnarray}
\omega_R \partial_t \omega_R &=& J\,\partial_t\,J + \Gamma\,\partial_t\,\Gamma\nonumber\\
\sqrt{J^2+\Gamma^2}(\partial_t+\Lambda)\sqrt{J^2+\Gamma^2}
&+&(\tilde\omega+\omega_R)(\partial_t+\Lambda)(\tilde\omega+\omega_R) = 0
\end{eqnarray}
for arbitrary $J(t)$ and $\Gamma(t)$,
where
\begin{equation}
\Lambda = -\frac{1}{2}\;\frac{\partial_t\left((\tilde\omega+\omega_R)^2+J^2+\Gamma^2\right)}{(\tilde\omega+\omega_R)^2+J^2+\Gamma^2}
\end{equation}
These relations are suggestive of a conformal nature to the transformation
and, with a little manipulation using the identities, one evaluates
\beq
\Gamma_{12}/\hbar=(\partial_t\theta) = 
\frac{(J\,\partial_t\,J + \Gamma\,\partial_t\,\Gamma)}{\omega_R\sqrt{J^2+\Gamma^2}}
\left[1 - \frac{(\tilde\omega+\omega_R)(\tilde\omega+2\omega_R)}{(\tilde\omega+\omega_R)^2+J^2+\Gamma^2}\right]
\label{connection}
\eeq
This quantity vanishes when $J^2+\Gamma^2$ is constant with respect to time.
Owing to the various identities, the result presented here can be expressed in many
equivalent forms.  In particular, it is zero when $\tilde\omega=0$.  The
equations of motion for the transformed fields are now
\beq
\left[
\begin{array}{cc}
 \hat L -\overline E_{12}+\hbar\omega_R & i\partial_t\theta\\
-i\partial_t\theta & \hat L -\overline E_{12}-\hbar\omega_R\\
\end{array}
\right]
\left(
\begin{array}{c}
\psi_+\\
\psi_-
\end{array}
\right) =0.
\eeq
In this basis, the centre of mass motion of the neutral atoms
factorizes from the wavefunction, since a neutral atom in an
electromagntic field is free on average. The two equations in the
matrix above may therefore be unravelled by introducing a
`gauge transformation' or `integrating factor'
\beq
\psi_\pm(x) = e^{\pm i \int_0^t X(t')dt'}\;\overline\psi(x),
\eeq
where the free wavefunction in $n=3$ dimensions is
\beq
\overline\psi(x) = \int \frac{d\omega}{(2\pi)}\frac{d^n{\bf k}}{(2\pi)^n}
\;e^{i({\bf k}\cdot{\bf x}-\omega t)}\delta \left(
\chi
\right)
\eeq
for the centre of mass motion is a general linear combination of plane
waves satisfying the dispersion relation for centre of mass motion
\beq
\chi = \frac{\hbar^2{\bf k}^2}{2m} + \hbar(\Omega-\omega) - \overline E_{12} = 0.
\eeq
The latter is enforced by the delta function.  This curious mixture of
continuous ($\omega$) and discontinuous ($\Omega$) belies the
effective nature of the model and the fact that its validity is only
for a continuous operation (an eternally sinusoidal radiation source
which never starts or stops).  The relevance of the model is thus
limited by this.  Substituting this form, we identify $X(t)$ as the
integrating factor for the uncoupled differential equations. The
complete solution is therefore
\beq
\psi_\pm(x) = e^{\mp i \int_0^t (\omega_R + i\partial_t\theta)dt'}\;\overline\psi(x).
\eeq
Notice that this result is an exact solution in the sense of being in
a closed form.  In the language of a gauge theory this result is gauge
dependent. This is because our original theory was not invariant under
time dependent transformations. The covariant procedure we have
applied is simply a method to transform the equations into an
appealing form; it does not imply invariance of the results under
a wide class of sources.

That this system undergoes transitions in time may be seen by
constructing wavefunctions which satisfy the boundary conditions where
the probability of being in one definite state of the system is zero
at $t=0$. To this end we write $\Psi_1 = \frac{1}{2}(\psi_++\psi_-)$ and
$\Psi_0 =  \frac{1}{2i}(\psi_+-\psi_-)$.
In order to proceed beyond this point it becomes necessary to
specify the initial value of $\Gamma_{12}$. This choice carries
with it physical consequences; the model is not invariant under
this choice. The obvious first choice is to set this to zero.
This would correspond to not making the rotating wave approximation
in the usual two level atom, with a cosine perturbation.
Focusing on the state $\Psi_0$ which was unoccupied at $t=0$ for $\Gamma_{12}=0$,
\beq
\Psi_0 = \sin\left(\int_0^t dt'\;\left[
\sqrt{\tilde\omega^2 +\hbar^{-2}J_0^2\cos^2(\Omega t')} -i\tilde\omega\frac{
J_0\Omega \sin(\Omega t')}{2\hbar\omega_R}\left[
\tilde\omega + \frac{J_0^2\cos^2(\Omega t')}{\hbar^2(\tilde\omega+\omega_R)}
\right]^{-1}
\right] \right)\overline\psi(x).
\eeq
We are interested in the period and amplitude of this quantity, whose
squared norm may be interpreted as the probability of finding the
system in the prepared state, given that it was not there at $t=0$.
Although the integral is then difficult to perform exactly, it is possible
to express it in terms of Jacobian elliptic integrals, logarithms and
trig functions. Nevertheless it is clear that
$\tilde\omega=\2(\tilde E_{21}/\hbar-\Omega)$ is the decisive
variable. When $\hbar\tilde\omega\ll J_0$ is small, the first
term is $J_0\cos(\Omega t)$ and the second term is small. This is
resonance, although the form of the solution is perhaps
unexpected. The form of the wavefunction guarantees a normalized result
which is regular at $\tilde\omega=0$ and one has
$\Psi_0 \sim \sin\left(
\int_0^t dt'\; \frac{J_0}{\hbar}\cos(\Omega t')
\right)$,
which may be compared to the standard result of the Jaynes-Cummings
model $\Psi_0 \sim \sin(J_0t/\hbar)$.  In the quantum case the
amplitude of the radiation source $J_0$ is quantized as an integral
number $N_\Omega$ of photons of frequency $\Omega$.  Here we see
modulation of the rate of oscillation by the photon frequency (or
equivalently the level spacing).  In a typical system the photon frequency
is several ten orders of magnitude larger than the coupling
strength $J_0 \ll
\hbar\Omega \sim \tilde E_{12}$ and thus there is an extremely
rapid modulation of the wavefunction. This results in an almost chaotic
collapse-revival behaviour with no discernable pattern, far from the
calm sinusiodal Rabi oscillations of the Jaynes-Cummings model.  If
$\hbar\tilde\omega\sim J_0$ the second term is of order unity and
then, defining the normalized resonant amplitude
\beq
A = \frac{J_0}{\sqrt{\hbar^2\tilde\omega^2+J_0^2}}
\eeq
one has
\beq
\Psi_0 \sim \sin\left(
\frac{J_0\Omega}{A}\;
E\left(\Omega t,A\right)
-A \,\int d(\Omega t)\;\frac{\sin(\Omega t)}{\sqrt{1 - A^2\sin^2(\Omega t)}}
\right)\overline\psi(x).
\eeq
The Jacobian elliptical integral $E(\alpha,\beta)$ is a doubly periodic function so one
could expect qualitatively different behaviour away from resonance.
On the other hand, far from resonance $\hbar \tilde\omega\gg
J_0$, the leading term of the connection becomes $\Psi_0 \sim \sin\left( \tilde\omega t
\right)\overline\psi(x) \sim\sin\left( \Omega t
\right)\overline\psi(x)$,
and the effect of the level spacing is washed out.

One can also consider other values for the connection.  Comparing
$\Gamma_{12}$ to the off diagonal sources $\gamma^\mu D_\mu$,
predicted on the basis of unitarity in effective non-equilibrium field
theory\cite{burgess12}, one obtains an indicatation that, if the
initial connection is in phase with the time derivative of the
perturbation, then one can effectively `resum' the decay processes
using the connection.  This is a backreaction effect of the time
dependent perturbation, or a renormalization in the language of
ref. \cite{burgess12}. If one chooses $\Gamma_{12}= J_0\sin(\Omega
t)$, this has the effect of making the off-diagonal terms in the
action not merely cosines but complex a conjugate pair $J_0\exp(\pm
i\Omega t)$. This corresponds to the result one obtains from making
the rotating wave approximation near resonance. This initial
configuation is extremely special.  With this choice, one has exactly
\beq
\Psi_0 = \sin\left(\int_0^t dt'\;\left[
\sqrt{\tilde\omega^2 +\hbar^{-2}J_0^2} 
\right] \right)\overline\psi(x).
\eeq
The stability of the solution is noteworthy, and the diagonalizing
transformation is rendered trivial.  The connection $\partial_t\theta$
is now zero under the diagonalizing transformation. Thus the above
result is exact and it is the standard result of the approximated
Jaynes-Cummings model. 
This indicates that the validity of the
Jaynes cummings model does not depend directly on its approximation,
but rather on the implicit choice of a connection.

The Jaynes-Cummings model is at the heart of many laser
phenomena. Korenman attempted to create a detailed microscopic
non-equilibrium field theory of the laser\cite{korenman1} from the
Lagrangian above; a recent letter\cite{henneberger1} proposes to do a
similar thing for semiconductor lasers.  In the first of these
Korenman claims to present a microscopic theory, but the dipole
interaction he considers is an effective interaction, as are the
variables for the pump part of the laser, so it is clearly not an {\em
ab initio} approach from the level of electron interactions.
Nevertheless, the structural elements, including exchange mechanisms
are present.  It is interesting to see how the model presented here is
related to the non-equilibrium field theory approach presented in
ref.\cite{burgess12}, which is directly related to the method used by
Korenman. There is a direct analogy between the two level system and
the forward and reverse time directions of the Schwinger closed time
path method, dictated by unitarity. Moreover the generalized model
presented in ref. \cite{burgess12} contains the present model in the
non-relativistic limit.

The form of the action in eqn (1) appears arbitrary but it may be
understood in the context of ref. \cite{burgess12} a conformal
perturbation. The action is perturbed by a time-dependent source which
one hopes would lead to a stable theory (the form of the action
remaining the same over time). This suggests conformal covariance. The
conformal link can be made by writing the above model as the
limit of a pseudo-relativistic theory since the conformal group is an
extension of the Lorentz group. This also makes contact with
ref. \cite{burgess12}.  The consistency of such an approach has been
verfied in ref.
\cite{ourpaper}. Beginning with the Lorentz covariant action
\beq
S = \int dV_t \left\{
\2 (\partial_\mu\phi_A)(\partial^\mu\phi_A) +\2 m_A\phi^2_A + \phi_A J_{AB}\phi_B
\right\},
\eeq
we consider a conformal rescaling by letting $g_{\mu\nu}\rightarrow
\Omega^2\;\overline g_{\mu\nu}$.  The action is not invariant under
this rescaling: if it were, there would be no need for the connection
$\Gamma$, or indeed this paper. The volume element scales as the
square root of the determinant of the metric,
i.e. $\sqrt{g}\rightarrow \Omega^4 \sqrt{\overline g}$ in $3+1$
dimensions, but we shall keep this separate for now. Since the issue
is not invariance but equivalence, this will not play a crucial
role. The first term in the braces contains one inverse power of the
metric, the second none and the third two. Choosing $\Omega^2 =
J_{AB}$, the off-diagonal, symmetric matrix with non-zero elements
$J(t)$, one can absorb the time dependent interaction by performing a
generalized rescaling. Rescaling the fields by
$\phi\rightarrow\Omega\overline\phi$, the action takes the form
\beq
S = \int dV_t' \left\{
\2 (D_\mu\overline\phi_A)(D^\mu\overline\phi_A) +\2 m'_A\overline\phi^2_A -\gamma^\mu_{AB} \overline
\phi(D_\mu\overline\phi)\right\},
\eeq
where $D^\mu = \partial^\mu\delta_{AB} + \gamma^\mu_{AB}$, which
is obtained by moving the scale factors through the derivatives, and
\beq
\2\frac{\partial_t \,J_{AB}}{J} = \left(\frac{\partial_t \Omega}{\Omega}\right) = \gamma_{AB}.
\eeq
The familar form of the conformal correction $\partial_\mu \Omega/\Omega$
is replaced in eqn. (1) simply by $\Gamma_{12} = \partial_t \Omega$, which
makes the initial value of $\Gamma_{12}\sim \partial_t J$ clear: it
is the connection required for the derivatives to commute with a
conformal rescaling brought about by the perturbation.

The non-relativistic limit of the transformed action with a time-only
dependent $\Omega(t)$ leads to eqn. (1), up to a Jacobian.  Thus
although these actions are not identical, they are related by an
overall factor which behaves as though to view the system through a
distorting glass. A similar non-equilibrium situation may be found in
ref. \cite{burgess10}. These two theories give essentially the same
results because they have the same structural elements. Indeed, it
might be possible to show that they form an equivalence class in the
formal sense, though this has not been shown here. It is this feature
of effective field theories which makes them robust and usable. The
result of this analysis is perhaps surprising though: the result which
one obtains by making the rotating wave approximation
$\Gamma_{12}\sim\sin(\Omega t)$ in a quantum mechanical formalism (the
JC model) is a stable, steady state effective field theory, but the
same model solved fully without approximation $\Gamma_{12}$ is not
stable to rescalings and leads to extremely complicated
behaviour. This is due as much to the method of analysis in time
dependent quantum mechanics as it is to do with the rotating wave
approximation. Here we have an example where such a model can be
solved exactly and does not appear to be give any qualitative pattern
of behaviour at experimental values typical for the
micromaser\cite{jaynes1}. What is interesting is why the specific
experimental situation should be well described by the very special
Jaynes-Cummings model rather than one of the many less stable theories
with a different initial $\Gamma_{12}$. This certainly warrants
further investigation, perhaps in the context of conformal field
theory.

I gratefully acknowledge Gabor Kunstatter and Meg Carrigton for their
friendly hospitality at Winnipeg in the autumn of 1996, where
most of the present work was done, and for invaluable discussions.
This work is supported by NATO collaborative research grant CRG950018.

%\bibliographystyle{unsrt}
%\bibliography{noneq}

\end{document}